\documentclass[aps,preprint,superscriptaddress]{revtex4}

\usepackage{epsfig,amsmath,amssymb}
\newcommand{\rv}{{\bf r}}

\newcommand{\T}{{\rm \sf t}}
\newcommand{\B}{{\sf B}}
\newcommand{\D}{{\sf D}}
\newcommand{\1}{{\bf 1}}
\renewcommand{\H}{{\sf H}}
\newcommand{\F}{{\sf F}}
\renewcommand{\P}{{\sf P}}
\renewcommand{\T}{{\sf T}}
\newcommand{\X}{{\sf X}}
\newcommand{\Y}{{\sf Y}}

\begin{document}

\title{Isometric and metamorphic operations on the space of local
  fundamental measures}

\author{Matthias Schmidt}
\affiliation{Theoretische Physik II, Physikalisches Institut, 
  Universit{\"a}t Bayreuth, D-95440 Bayreuth, Germany}
\affiliation{H. H. Wills Physics Laboratory, University of Bristol, 
  Royal Fort, Tyndall Avenue, Bristol BS8 1TL, United Kingdom}

\date{15 November 2010, revised: 23 December 2010, to appear in Molecular Physics}

\begin{abstract}
   We consider symmetry operations on the four-dimensional vector
   space that is spanned by the local versions of the Minkowski
   functionals (or fundamental measures): volume, surface, integral
   mean curvature, and Euler characteristic, of an underlying
   three-dimensional geometry.  A bilinear combination of the measures
   is used as a (pseudo) metric with $++--$ signature, represented by
   a $4\times 4$ matrix with unit entries on the counter diagonal. Six
   different types of linear automorphisms are shown to leave the
   metric invariant. Their generators form a Lie algebra that can be
   grouped into two mutually commuting triples with non-trivial
   structure constants. We supplement these six isometric operations
   by further ten transformations that have a metamorphic (altering)
   effect on the underlying geometry. When grouped together, four
   different linear combinations of the metamorphic generators form a
   previously obtained third-rank tensor. This is shown to describe
   four different types of mutually commuting ``shifting'' operations
   in fundamental measure space.  The relevance for fundamental
   measures density functional theory is discussed briefly.
\end{abstract}

\maketitle

\section{Introduction}

Applying classical density functional theory (DFT) requires to have an
approximation for the Helmholtz free energy as a functional of the
one-body density distribution(s) \cite{evans79,evans92}.  For the case
of additive hard sphere mixtures, Rosenfeld's fundamental measures
theory (FMT) \cite{rosenfeld89} is an approximate DFT that unified
several earlier liquid state theories, including the Percus-Yevick
integral equation theory and scaled-particle theory, and encapsulates
their results in a free energy functional. Several recent reviews give
a detailed account of FMT and some of its extensions and modifications
\cite{tarazona08review,roth10review,evans10review}. The theory was
used to address a broad variety of interesting equilibrum phenomena,
ranging from freezing to capillary behaviour of liquids. When compared
to computer simulation data, theoretical results for e.g.\ density
profiles and interfacial tension were typically found to be very
reliable.  FMT rests on building weighted densities via convolution
with the bare density profile(s).  The microscopic density profile
$\rho_i(\rv)$ of species $i$ gives the mean number of particles of
species $i$ in an infinitesimal volume element at given position~$\rv$
and hence carries dimensions of (length)$^{-3}$. The weighted
densities in FMT are smoothed versions of these ``real'' density
distributions.  In Kierlik and Rosinberg's (KR) version
\cite{kierlik90} of FMT \cite{rosenfeld89}, there are four scalar
weight functions for each hard sphere species.  Rosenfeld's original
approach that involves additional weight functions was shown to be
equivalent to the KR version \cite{phan93}, and was later carried much
further by Tarazona \cite{tarazona00} and Cuesta et al
\cite{cuesta02}. FMT has intimate connections to methods from integral
geometry \cite{mecke} via the Gauss-Bonnet theorem
\cite{rosenfeldGaussBonnet}.

The weight functions in FMT are quantities with dimension of negative
integer powers of length, ranging from $({\rm length})^{-3}$ to $({\rm
  length})^0$. A linear combination of pairs of weight functions that
are convolved with each other is used to express the Mayer bond
$f_{ij}(r)$, as a function of distance $r$. Recall that for hard
sphere mixtures, the Mayer bond equals $f_{ij}(r)=-1$ for distances
$r<R_i+R_j$, i.e.\ when the two spheres with radii $R_i$ and $R_i$
overlap, and it vanishes otherwise. Here the subscripts $i,j$ label
the different species.  Originally proposed for hard spheres, this
framework was sufficient to to derive FMTs for models such as the
Asakura-Oosawa colloid-polymer mixture \cite{schmidt00cip} and the
Widom-Rowlinson model \cite{schmidt01wr}. However, the treatment of
binary non-additive hard sphere mixtures required significant
modification of the mathematical structure of FMT
\cite{schmidt04nahs}. In particular, further weight functions were
introduced in order to correctly model the deviation of the hard core
interaction range between species $i$ and $j$ from the sum of their
radii, $R_i+R_j$. The fact that this deviation is non-vanishing is the
defining feature of non-additive hard sphere mixtures. The additional
weight (or kernel) functions possess dimensionalities up to
(length)$^{-6}$. They can be grouped in a double-indexed tensorial
form \cite{schmidt04nahs} and were shown to possess a remarkable group
structure \cite{schmidt07nage}. In very recent work, the FMT for
non-additive hard spheres was applied successfully to bulk structure
\cite{ayadim10,hopkins10nahs} and to interfacial phenomena
\cite{hopkins10nahs}.

Several features of the mathematics that underlies the FMT weight
functions have emerged \cite{schmidt04nahs,schmidt07nage}: i) The four
different position-dependent fundamental measures (in the KR
formulation) can be viewed as elements of an abstract four-dimensional
vector space. ii) Based on dimensional analysis, a (pseudo) metric can
be defined, which can be represented by a 4$\times$4 matrix with unit
entries on the counter-diagonal. All other entries in this matrix
vanish.  The metric has a $++-\,-$ signature, hence it differs both
from that of Minkowski spacetime in special relativity ($+--\,-$) and
from that of four-dimensional Euclidian space ($+++\,+$). iii)
Operations that are common in linear algebra, i.e.\ matrix
multiplication and more general contraction of tensor indices possess
meaningful interpretation, see e.g.\ the shifting transform described
in Ref.\ \cite{schmidt07nage}. Here all product operations are carried
out in Fourier space and hence correspond to convolutions in real
space.

In the present paper we explore the mathematical structure further by
focusing on symmetry operations that leave the fundamental measure
metric invariant. Our motivation comes from the fact that careful
analysis of the symmetries is central to exploiting the properties of
any (abstract) space. Typically, this tasks requires the
identification of the linear automorphisms that leave the metric
invariant. Recall that an automorphism is a bijective function that
maps a space onto itself (i.e.\ both function value and argument are
element of the same space).  Much structure can be revealed by
considering infinitesimal versions, or generators, of the
transformations. In a {\em Lie algebra} the commutator of any pair of
generators can be represented as a linear combination of again the
same generators. The coefficients of the linear combinations form the
structure constants of the algebra.

In Euclidian space, the symmetry operations that leave the metric
invariant are orthogonal transformations, or rotations. These possess
three (six) independent generators in three (four) spatial
dimensions. For the case of Minkowski spacetime with three spatial and
one time-like dimension, there are three spatial rotations and three
Lorentz transformations, or boosts, the latter coupling time and one
of the spatial dimensions. The number of degrees of freedom, and hence
the dimensionality of the group of isometries, is independent of the
signature of the metric. However, the algebraic structure, as
expressed by commutator relations between the respective
(infinitesimal) generators of the transforms, differs for both cases.
For the case of spacetime, the resulting mathematical structure is the
Lorentz group. (One refers to the Poincar\'e group when four
translations in the different spacetime directions are added.) Here we
present in detail a similar analysis for the space of Minkowski
functionals \cite{mecke}. We describe four boosts and two rotations
that leave the metric invariant. These are complemented by further ten
operations that change the metric and that we refer to as metamorphic
operations.  We show that the spherical shifting operation of
Ref.\ \cite{schmidt07nage} is readily generalized to four different
types of shifting, and that the corresponding generators can be
expressed as linear combinations of the metamorphic generators.

The paper is organized as follows. In Sec.\ \ref{sec:theory} the
theory is laid out, including the description of inner boosts and
inner rotations as isometric transformations
(Sec.\ \ref{sec:boostsAndRotations}), of metamorphic operations
(Sec.\ \ref{sec:deformations}), and the relationship of Jeffrey's
third-rank tensor \cite{schmidt07nage} to the latter
(Sec.\ \ref{sec:translations}). Concluding remarks are given in
Sec.\ \ref{sec:conclusions}.

\section{Transforming the fundamental measures}
\label{sec:theory}
\subsection{Metric and inner scalar product}
We consider a four-dimensional real vector space with elements ${\sf
  u}=(u_0,u_1,u_2,u_3)$, where $\sf u$ depends on the
three-dimensional argument $\bf q$ in Fourier space. The dependence on
three-dimensional position $\rv$ is then obtained by inverse Fourier
transform, $(2\pi)^{-3}\int d{\bf q} e^{i {\bf q}\cdot \rv} {\sf u}$.
The vector components $u_\nu$, with index $\nu=0,1,2,3$, are
dimensional objects: $u_\nu$ possesses the dimension
(length)$^\nu$. Hence $u_3$ is a measure of volume, $u_2$ of surface,
$u_1$ of mean curvature, and $u_0$ of Gaussian curvature. We let the
$u_\nu$ take on arbitrary (real) values, and hence restrict ourselves
not to cases where the measures describe an underlying geometrical
body. The interpretation of the $u_\nu$ in terms of geometric measures
is only intended to guide the intuition, the mathematics that we
present in the following is based on formal arguments.

We use the metric represented by the matrix
\begin{align}
  {\sf M} =
  \left(\begin{matrix}
    0 & 0 & 0 & 1 \\
    0 & 0 & 1 & 0 \\
    0 & 1 & 0 & 0 \\
    1 & 0 & 0 & 0
  \end{matrix}\right),
  \label{eq:metric}
\end{align}
hence a measure of squared ``length'' of a vector $\sf u$ is given by
${\sf u}^{\sf t}\cdot {\sf M}\cdot{\sf u}=2({\sf u}_0{\sf u}_3+{\sf
  u}_1{\sf u}_2)$, where the superscript $\sf t$ indicates matrix
transposition, and the dot indicates matrix multiplication. The scalar
product between two vectors $\sf u$ and $\sf v$ is ${\sf u}^{\sf
  t}\cdot{\sf M}\cdot{\sf v}={\sf u}_0{\sf v}_3+{\sf u}_1{\sf
  v}_2+{\sf u}_2{\sf v}_1+{\sf u}_3{\sf v}_0$.  Clearly, this is
symmetric upon interchange of the vectors, i.e.\ ${\sf u}^{\sf
  t}\cdot{\sf M}\cdot{\sf v}={\sf v}^{\sf t}\cdot{\sf M}\cdot{\sf u}$.
The eigenvalues of ${\sf M}$ are $-1$ and $1$, both are doubly
degenerate; hence $\sf M$ possesses $(++--)$ signature.  As $\sf M$ is
not positive definite (i.e.\ not all of its eigenvalues are positive),
it can yield negative squared distances and hence constitutes not a
metric in the strict sense, but one refers to a pseudo metric. While
it is enirely possible to discriminate between covariant and
contravariant vectors and correspondingly introduce lower and upper
indices, which can be interchanged by application of the metric, we
will not do so in the following. The present paper is primarily
concerned with second-rank tensors, and we find the (index-free)
matrix notation to be simpler, and will primarily rely on this in what
follows.

The hard sphere weight functions of FMT can serve as an example. These
are functions of the squared wave number $q^2$ and the radius $R$ of
the hard spheres. The Fourier space expressions of the KR version of
the weight functions are $w_0=c+(qRs/2)$, $w_1=(qRc+s)/(2q)$,
$w_2=4\pi Rs/q$, $w_3=4\pi(s-qRc)/q^3$, where $s=\sin(qR)$ and
$c=\cos(qR)$. The real space expression that corresponds, via inverse
Fourier transform, to $w_3$ is a unit step function with range of $R$,
i.e.\ $\Theta(R-|\rv|)$, where $\Theta(\cdot)$ is the Heaviside (step)
function. Within our framework we view the $w_\nu$ as the four
components of a vector $\sf w$.  By straightforward explicit algebra
one can show that ${\sf w}^{\sf t}\cdot{\sf M}\cdot {\sf w}$ yields
the Fourier transform of a unit step function with range $2R$,
i.e.\ the expression for $w_3$ given above, but with $R$ replaced by
$2R$.  Explicitly this is
$2(w_0w_3+w_1w_2)=4\pi[\sin(2qR)-2qR\cos(2qR)]/q^3$. The significance
in statistical physics stems from the fact this is (up to a trivial
minus sign) the Fourier transform of the negative Mayer function of
the pair potential of hard spheres of radius $R$.  For a mixture, the
additional species possesses weight functions $v_\nu$ of range $R'$,
given by the above expressions for $w_\nu$, but with $R$ being
replaced by $R'$. It is straightforward to verify that ${\sf
  v}\cdot{\sf M}\cdot {\sf
  w}=4\pi[\sin(q(R+R'))-q(R+R')\cos(q(R+R'))]/q^3$, which again is the
above expression for the unit step, $w_3$, but with $R$ replaced by
the sum of the radii, $R+R'$. These identities constitute one of the
central building blocks of KR's formulation of FMT. The generalization
to non-additive mixtures \cite{schmidt04nahs} amounts to introducting
$4\times 4$ matrices that change the range of the weight functions
$w_\nu$.  This shifting operation is discussed in detail in
Ref.\ \cite{schmidt07nage}. Below in Sec.\ \ref{sec:translations} we
give further three such ``internal'' shifting operations. We emphasize
that all transformations that are considered here are of internal
nature, i.e.\ act on the four-dimensional space of fundamental
measures, as opposed to e.g.\ translations and rotations of the
underlying three-dimensional Euclidian space, which we do not consider
here.

The central aim of this paper is to formulate linear automorphisms
that leave the metric (\ref{eq:metric}) invariant. We refere to such
operations as isometries on the space of fundamental measures. Hence
one has to identify $4\times 4$ transformation matrices $\sf A$ that
obey
\begin{align}
  {\sf A}^{\sf t} \cdot {\sf M} \cdot {\sf A} = 
  {\sf A} \cdot {\sf M} \cdot {\sf A}^{\sf t} = {\sf M},
  \label{eq:invariance}
\end{align}
which implies that a vector $\sf u$ and its transform ${\sf
  A}\cdot{\sf u}$ possess the same squared modulus. This can be seen
from $({\sf A}\cdot {\sf u})^{\sf t}\cdot {\sf M}\cdot ({\sf A}\cdot
{\sf u})={\sf u}^{\sf t}\cdot {\sf A}^{\sf t}\cdot {\sf M}\cdot {\sf
  A}\cdot {\sf u} = {\sf u}^{\sf t}\cdot {\sf M}\cdot {\sf u}$, where
the last equality follows from (\ref{eq:invariance}). An alternative
is obtained by multiplying (\ref{eq:invariance}) from the right by the
inverse ${\sf A}^{-1}$, and from the left by $\sf M$, and observing
that ${\sf M}^2=\1$, where $\1$ is the $4\times 4$ unit matrix. Hence
\begin{equation}
  {\sf M}\cdot {\sf A}^{\sf t} \cdot {\sf M} = {\sf A}^{-1}.
  \label{eq:inverse}
\end{equation}
Note that this differs from the condition for orthogonal matrices,
${\sf A}^{\sf t}={\sf A}^{-1}$. While transposition can be viewed as
mirroring the matrix elements on the diagonal, the operation on the
left hand side of (\ref{eq:inverse}) corresponds to mirroring the
matrix elements on the counter diagonal.

\subsection{Inner rotations and boosts as isometries}
\label{sec:boostsAndRotations}

Let us formulate the linear isometries, i.e.\ the automorphisms $\sf
A$ that obey (\ref{eq:invariance}), by choosing appropriate generators
$\sf X$ for each different type of transform, where $\sf X$ is a
$4\times 4$ matrix. The transformation matrices $\sf A$ are then
obtained by (matrix) exponentiation. In order to see this, consider
that the expression $\1+{\sf X}d\tau$ can be viewed as an
infinitesimal transform of differential magnitude $d\tau$.  A
transform by a finite amount $\tau$ can then obtained in the continuum
limit of $N$-fold application of the infinitesimal transform, where
each step is taken to be of magnitude $\tau/N$. This amounts to
$\lim_{N\to\infty}(\1+\tau{\sf X}/N)^N=\exp(\tau{\sf X})\equiv {\sf
  A}$, where the result depends parametrically on $\tau$ and the form
of $\sf A$ is specific to that of $\sf X$. Here the exponential of a
matrix is defined by its power series $\exp(\tau {\sf
  X})=\sum_{m=0}^\infty(\tau^m/m!){\sf X}^m$. In the following, we
allow the transformation parameter $\tau$ to be dimensional, i.e.\ to
carry a non-vanishing power of length scale.

In order to allow for meaningful matrix multiplication (as is
necessary for matrix exponentiation) the generators need to possess
matrix components with suitable dimensionalities. This implies that
the product of the transformation parameter and a matrix entry, $\tau
{\sf X}_{\mu\nu}$, where $\mu$ enumerates the rows and $\nu$
enumerates the columns, with both indices running from 0 to 3, must be
of unit (length)$^{\mu-\nu}$.  Taking matrix powers then preserves the
ordering of dimensions, i.e.\ the $\mu\nu$-component of the $m$-th
matrix power, $(\tau^m {\sf X}^m)_{\mu\nu}$, has the same
dimensionality as $\tau {\sf X}_{\mu\nu}$ itself. Hence we can
exponentiate the generators and obtain finite transforms. Besides
letting $\tau$ be a dimensional object, in the following the only
further dependence on length scale shall be via $q^2$, the squared
argument in Fourier space. This corresponds to the (negative)
Laplacian in the corresponding real three-dimensional space.

From general arguments for four-dimensional spaces, we expect the
isometry group to be six-dimensional, i.e.\ to possess six linearly
independent generators, cf.\ the cases of Euclidian space and
Minkowski spacetime of special relativity mentioned above.  Given a
set of such generators, $\{{\sf X}_\alpha\}$, ennumerated by index
$\alpha$, a general transform is obtained as $\exp(\sum_\alpha
\tau_\alpha {\sf X}_\alpha)$, where $\tau_\alpha$ is the magnitude of
the $\alpha$-th transform.  In principle the different contributions
to the the total transform can be disentangled via the
Baker-Campbell-Hausdorff formula. This requires knowledge of the
algebraic group structure, which is encoded in commutator relations
between the different generators, as laid out below.

Here we discriminate between four generators for boosts, ${\sf
  B}_\alpha$, and two generators for rotations, ${\sf D}_\alpha$.  The
subscript indicates the dimensionality; the $\mu\nu$-element of a
given generator matrix possesses units of (length)$^{\mu-\nu-\alpha}$.
As laid out above, all elements along a given diagonal possess the
same dimensionality; the dimensionality then decreases (increases) by
one power of length scale when moving up (down) to the next diagonal.
We call boosts those generators that satisfy ${\sf B}_\alpha\cdot{\sf
  B}_\alpha=q^{2\alpha}\1$. Generator of rotations are those that
satisfy ${\sf D}_\alpha\cdot{\sf D}_\alpha=-q^{2\alpha}\1$. The
generators are not unique; one can always build linear combinations to
obtain a different formulation. Here we choose the following set of
generators.
\begin{align}
  \B_0 = 
  \left(\begin{matrix}
    1 & 0 & 0 & 0 \\
    0 & 1 & 0 & 0 \\
    0 & 0 & -1 & 0 \\
    0 & 0 & 0 & -1
  \end{matrix}\right),\quad
  \B_2 = 
  \left(\begin{matrix}
    0 & 0 & -q^4 & 0 \\
    0 & 0 & 0 & q^4 \\
    -1 & 0 & 0 & 0 \\
    0 & 1 & 0 & 0
  \end{matrix}\right),\quad
  \D_2 =
  \left(\begin{matrix}
    0 & 0 & -q^4 & 0 \\
    0 & 0 & 0 & q^4 \\
    1 & 0 & 0 & 0 \\
    0 & -1 & 0 & 0
  \end{matrix}\right),\label{eq:automorphismGenerators1}\\
  \B_{0'} = 
  \left(\begin{matrix}
    1 & 0 & 0 & 0 \\
    0 & -1 & 0 & 0 \\
    0 & 0 & 1 & 0 \\
    0 & 0 & 0 & -1
  \end{matrix}\right),\quad
  \B_1 = 
  \left(\begin{matrix}
    0 & -q^2 & 0 & 0 \\
    -1 & 0 & 0 & 0 \\
    0 & 0 & 0 & q^2 \\
    0 & 0 & 1 & 0
  \end{matrix}\right),\quad
  \D_1 =
  \left(\begin{matrix}
    0 & -q^2 & 0 & 0 \\
    1 & 0 & 0 & 0 \\
    0 & 0 & 0 & q^2 \\
    0 & 0 & -1 & 0
  \end{matrix}\right),
  \label{eq:automorphismGenerators2}
\end{align}
We have grouped the generators into two families, each consisting of
two boosts and one rotation. The first one consists of
$\B_0,\B_2,\D_2$ and is given in (\ref{eq:automorphismGenerators1}),
the second one consists of $\B_{0'},\B_1,\D_1$ and is given in
(\ref{eq:automorphismGenerators2}).  Both families form closed Lie
algebras, constituted by the commutator relations
\begin{align}
 [\B_0,\B_2]&=2\D_2,\quad [\B_0,\D_2]=2\B_2,\quad [\B_2,\D_2]=2q^4\B_0,\\
 [\B_{0'},\B_1]&=2\D_1,\quad [\B_{0'},\D_1]=2\B_1,\quad [\B_1,\D_1]=2q^2\B_{0'},
\end{align}
where the commutator between two matrices $\X$ and $\Y$ is defined as
$[\X,\Y]=\X\cdot\Y-\Y\cdot\X$. Members of different families commute;
these are pairs of boosts:
$[\B_0,\B_{0'}]=[\B_0,\B_1]=[\B_{0'},\B_2]=[\B_1,\B_2]=0$, the (only)
pair of rotations: $[\D_1,\D_2]=0$, and the four mixed pairs of a
rotation and a boost:
$[\B_0,\D_1]=[\B_1,\D_2]=[\B_2,\D_1]=[\B_{0'},\D_2]=0$.
Tab.\ \ref{tab:automorphismAlgebra} gives an overview of the group
structure in table format. All relationships can be obtained by
straightforward matrix algebra. We give a full multiplication table in
Tab.\ \ref{tab:automorphismMultiplication}; anti-commutator relations
are included for completeness. Note that in each family already the
bare products (not commutators) give the result of the commutators up
to a factor of 2. As a consequence, the anti-commutators within each
sub-algebra vanish, see Tab.\ \ref{tab:automorphismAlgebra}. Altough
commutators between members of different sub-algebras vanish, their
plain products do not,
cf.\ Tab.\ \ref{tab:automorphismMultiplication}. The nine matrices
that result from the products (as referred to in the off-diagonal
blocks in Tab.\ \ref{tab:automorphismMultiplication}) will be used
below in order to define further, metamorphic, operations on the
fundamental measures.

\begin{table}[h]
\begin{equation*}
\begin{array}{|c||ccc|ccc|}\hline
[\X,\Y]/2& \B_0 & \B_2 & \D_2 & \B_{0'} & \B_1 & \D_1 \\
\hline\hline
  \B_0 &  0 & \D_2 & \B_2 &0&0&0\\
  \B_2 & -\D_2 & 0 & q^4\B_0 &0&0&0\\
  \D_2 & -\B_2 & -q^4\B_0 & 0 &0&0&0\\
\hline
  \B_{0'}& 0 & 0 & 0 & 0 & \D_1 & \B_1\\
  \B_1   & 0 & 0 & 0 & -\D_1 & 0 & q^2\B_{0'} \\
  \D_1   & 0 & 0 & 0 & -\B_1 & -q^2\B_{0'} & 0 \\\hline
\end{array}
\end{equation*}
\caption{Table for commutator relationships $[{\sf X},{\sf Y}]/2$ for
  the generators of boosts, $\B_\nu$, and rotations, $\D_\nu$. $\X$
  denotes a matrix of the leftmost column, $\Y$ one of the top row.}
\label{tab:automorphismAlgebra}
\end{table}

\begin{table}[h]
\begin{equation*}
\begin{array}{|c||ccc|ccc|}\hline
\X\cdot\Y& \B_0 & \B_2 & \D_2 & \B_{0'} & \B_1 & \D_1 \\
\hline\hline
  \B_0 & \1 & \D_2 & \B_2 & \P_0 & \H_1 & \F_1 \\
  \B_2 & -\D_2 & q^4\1 & q^4\B_0 & \H_2 & \F_3& \P_3\\
  \D_2 & -\B_2 & -q^4\B_0 & -q^4\1 & \F_2 & \P_{3'} & \F_{3'} \\
\hline
  \B_{0'}& \P_0 & \H_2 & \F_2 & \1 & \D_1 & \B_1\\
  \B_1   & \H_1 & \F_3& \P_{3'} & -\D_1 & q^2\1 & q^2\B_{0'} \\
  \D_1   & \F_1 & \P_3& \F_{3'} & -\B_1 & -q^2\B_{0'} & -q^2\1 \\\hline
\end{array}
\quad
\begin{array}{|c||ccc|ccc|}\hline
\{\X,\Y\}/2& \B_0 & \B_2 & \D_2 & \B_{0'} & \B_1 & \D_1 \\
\hline\hline
  \B_0 & \1 & 0 & 0 & \P_0 & \H_1 & \F_1 \\
  \B_2 & 0 & q^4\1 & 0 & \H_2 & \F_3& \P_3\\
  \D_2 & 0 & 0 & -q^4\1 & \F_2 & \P_{3'} & \F_{3'} \\
\hline
  \B_{0'}& \P_0& \H_2& \F_2 & \1 & 0 & 0\\
  \B_1   & \H_1& \F_3& \P_{3'} & 0 & q^2\1 & 0 \\
  \D_1   & \F_1& \P_3& \F_{3'} & 0 & 0 & -q^2\1 \\\hline
\end{array}
\end{equation*}
\caption{Left: Multiplication table $\X\cdot\Y$ for products of the
  generators of boosts and rotations.  Right: Table of anti-commutator
  relationships $\{{\sf X},{\sf Y}\}/2$ for the generators of boosts
  and rotations. In both tables $\X$ denotes a matrix of the leftmost
  column, $\Y$ one of the top row. }
\label{tab:automorphismMultiplication}
\end{table}

It is now straightforward to calculate finite transforms via
exponentiation of the respective generators multiplied by its
transformation parameter. Recall that the latter is a dimensional
object, and that the most general finite transform is given by
$\exp(\sum_\nu \tau_\nu {\sf X}_\nu)$, where $\tau_\nu$ possess
dimensions of (length)$^\nu$. Here we give only the results for the
case where all parameters bar one vanish. These are the following
expressions for finite transformations corresponding
to~(\ref{eq:automorphismGenerators1})
\begin{align}
\exp(\tau_0\B_0)=&  \left(\begin{matrix}
    e^{\tau_0} & 0 & 0 & 0\\
    0 & e^{\tau_0} & 0 & 0\\
    0 & 0 & e^{-\tau_0} & 0\\
    0 & 0 & 0 & e^{-\tau_0}
  \end{matrix}
  \right),\label{eq:finiteB0}\\
  \exp(\tau_2\B_2)=&  \left(\begin{matrix}
    \cosh(\tau_2q^2) & 0 & -q^2\sinh(\tau_2q^2) & 0\\
    0 & \cosh(\tau_2q^2) & 0 & q^2\sinh(\tau_2q^2)\\
    -q^{-2}\sinh(\tau_2q^2) & 0 & \cosh(\tau_2q^2) & 0\\
    0 & q^2\sinh(\tau_2q^2) & 0 & \cosh(\tau_2q^2)
  \end{matrix}
  \right),\label{eq:finiteB2}\\
\exp(\tau_2\D_2)=&  \left(\begin{matrix}
    \cos(\tau_2q^2) & 0 & -q^2\sin(\tau_2q^2) & 0\\
    0 & \cos(\tau_2q^2) & 0 & q^2\sin(\tau_2q^2)\\
    q^{-2}\sin(\tau_2q^2) & 0 & \cos(\tau_2q^2) & 0\\
    0 & -q^{-2}\sin(\tau_2q^2)& 0 & \cos(\tau_2q^2)
  \end{matrix}
  \right).\label{eq:finiteD2}
\end{align}
When applied to a vector $\sf u$, (\ref{eq:finiteB0}) describes a
multiplication of the components $u_0$ and $u_1$ by $e^{\tau_0}$, and
division of $u_2$ and $u_3$ by the same constant. Trivially, the
(pseudo) squared modulus $2(u_1u_2+u_0u_3)$ is left
unchanged. Eq.~(\ref{eq:finiteB2}) is reminiscent of a hyperbolic
rotation, and (\ref{eq:finiteD2}) of an ordinary rotation. Note the
difference in occurrence of the minus signs in (\ref{eq:finiteB2}) and
(\ref{eq:finiteD2}).

For the generators (\ref{eq:automorphismGenerators2}) we obtain the
following finite transforms:
\begin{align}
\exp(\tau_0\B_{0'})=&  \left(\begin{matrix}
    e^{\tau_0} & 0 & 0 & 0\\
    0 & e^{-\tau_0} & 0 & 0\\
    0 & 0 & e^{\tau_0} & 0\\
    0 & 0 & 0 & e^{-\tau_0}
  \end{matrix}
  \right),\label{eq:finiteB0s}\\
\exp(\tau_1\B_1)=&  \left(\begin{matrix}
    \cosh(\tau_1q) & -q\sinh(\tau_1q) & 0 & 0\\
    -q^{-1}\sinh(\tau_1q) & \cosh(\tau_1q) & 0 & 0\\
    0 & 0 & \cosh(\tau_1q) & q\sinh(\tau_1q)\\
    0 & 0 & q^{-1}\sinh(\tau_1q) & \cosh(\tau_1q)
  \end{matrix}
  \right),\label{eq:finiteB1}\\
\exp(\tau_1\D_1)=&  \left(\begin{matrix}
    \cos(\tau_1q) & -q\sin(\tau_1q) & 0 & 0\\
    q^{-1}\sin(\tau_1q) & \cos(\tau_1q) & 0 & 0\\
    0 & 0 & \cos(\tau_1q) & q\sin(\tau_1q)\\
    0 & 0 & -q^{-1}\sin(\tau_1q) & \cos(\tau_1q)
  \end{matrix}
  \right),\label{eq:finiteD1}
\end{align}
Again (\ref{eq:finiteB0s}) induces a straightforward scaling of vector
components, (\ref{eq:finiteB1}) is a hyperbolic rotation and is
(\ref{eq:finiteD1}) an ordinary rotation. Recall that hyperbolic
rotation can be viewed as Lorentz transforms (and vice versa).

As a summary, we have identified six real matrices
$\B_0,\B_{0'},\B_1,\B_2,\D_1$, and $\D_2$, that posses the algebraic
structure shown in Tab.\ \ref{tab:automorphismMultiplication}. The
general (real) linear group, i.e.\ that of all real $4\times 4$
matrices, is $4^2=16$ dimensional. Besides the unit matrix, this
leaves nine matrices to be considered. In the following we will use
the matrices obtained as products of two isometric generators,
cf.\ Tab.\ \ref{tab:automorphismMultiplication}. We find it
interesting to investigate their action, when viewed as infinitesimal
transformations, on the space of fundamental measures. Clearly, they
cannot generate isometries -- we have exhausted these already. Hence
we expect that the metric will not be conserved under the application
of these further transformations, and we will henceforth refer to
these transformations as metamorphic, as they change the underlying
geometry in a fundamental way.

The difference between automorphism and metamorphisms is reflected in
the symmetry properties of their generators. The isometric generators
(\ref{eq:automorphismGenerators1}) and
(\ref{eq:automorphismGenerators2}) are anti-symmetric with respect to
mirroring on the counterdiagonal, i.e. each generator $\X$ satisfies
\begin{align}
  {\sf M} \cdot \X^{\sf t} \cdot {\sf M} = -\X.
  \label{eq:automorphismGeneratorSymmetry}
\end{align}
This can be seen by inserting the infinitesimal versions ${\sf A}=1+\X
d\tau$ and ${\sf A}^{-1}=1-\X d\tau$ into (\ref{eq:inverse}). Note
that the symmetry (\ref{eq:automorphismGeneratorSymmetry}) leaves 6
parameters free, which is consistent with the dimensionality of the
corresponding group of transformations (and hence the number of
generators).  Correspondingly, metamorphic generators are symmetric
under mirroring on the counter-diagonal, i.e.\ they satisfy
\begin{align}
  {\sf M} \cdot \X^{\sf t} \cdot {\sf M} = \X,  
  \label{eq:metamorphismGeneratorSymmetry}
\end{align}
as we will see in the following. Note that the symmetry
(\ref{eq:metamorphismGeneratorSymmetry}) leaves 10 parameters
undetermined.

\subsection{Metamorphic transformations}
\label{sec:deformations}

We start by giving the explicit expressions for the matrices that we
choose as generators of the metamorphic operations. As above, the
index $\nu$ of a given generator $\X_\nu$ indicates its
dimensionality. Explicit expressions for the nine different generators
are as follows.
\begin{align}
\F_1=\left(\begin{matrix}
    0 & -q^2 & 0 & 0\\
    1 & 0 & 0 & 0\\
    0 & 0 & 0 & -q^2\\
    0 & 0 & 1 & 0
  \end{matrix}
  \right),\quad
&\F_2=\left(\begin{matrix}
    0 & 0 & -q^4 & 0\\
    0 & 0 & 0 & -q^4\\
    1 & 0 & 0 & 0\\
    0 & 1 & 0 & 0
  \end{matrix}
  \right),\quad
\F_3=\left(\begin{matrix}
    0 & 0 & 0 & -q^6\\
    0 & 0 & q^4 & 0\\
    0 & q^2 & 0 & 0\\
    -1 & 0 & 0 & 0
  \end{matrix}
  \right),\label{EQdeformation1}
\end{align}
\begin{align}
\H_1=\left(\begin{matrix}
    0 & -q^2 & 0 & 0\\
    -1 & 0 & 0 & 0\\
    0 & 0 & 0 & -q^2\\
    0 & 0 & -1 & 0
  \end{matrix}
  \right),\quad
&\H_2=\left(\begin{matrix}
    0 & 0 & -q^4 & 0\\
    0 & 0 & 0 & -q^4\\
    -1 & 0 & 0 & 0\\
    0 & -1 & 0 & 0
  \end{matrix}
  \right),\quad
\F_{3'}=\left(\begin{matrix}
    0 & 0 & 0 & -q^6\\
    0 & 0 & -q^4 & 0\\
    0 & -q^2 & 0 & 0\\
    -1 & 0 & 0 & 0
   \end{matrix}\right), \label{EQdeformation2}
\end{align}
\begin{align}
\P_0=\left(\begin{matrix}
    1 & 0 & 0 & 0\\
    0 & -1 & 0 & 0\\
    0 & 0 & -1 & 0\\
    0 & 0 & 0 & 1
   \end{matrix}\right),\quad
&\P_3=\left(\begin{matrix}
    0 & 0 & 0 & -q^6\\
    0 & 0 & -q^4 & 0\\
    0 & q^2 & 0 & 0\\
    1 & 0 & 0 & 0
   \end{matrix}\right),\quad
\P_{3'}=\left(\begin{matrix}
    0 & 0 & 0 & -q^6\\
    0 & 0 & q^4 & 0\\
    0 & -q^2 & 0 & 0\\
    1 & 0 & 0 & 0
   \end{matrix}\right). \label{EQdeformation3}
\end{align}
Here we have grouped the nine generatores into three Abelian
subgroups, given in (\ref{EQdeformation1}), (\ref{EQdeformation2}),
and (\ref{EQdeformation3}), respectively. Any pair of matrices from of
one of these subgroups satisfies $[\X_\mu,\Y_\nu]=0$.  In general, the
commutator between matrices from different subgroups is (up to a minus
sign) a multiple of $q^2$ times an isometric generator.  Some of these
pairs commute. The complete algebra of commutator relationships
between the metamorphic generators is summarized in
Tab.\ \ref{TABcommutatorsOfMetamorphisms}.  Remarkably, the commutator
between any two pairs of these matrices either vanishes or it is a
multiple of one of the isometric generators of
Sec.\ \ref{sec:boostsAndRotations}. Inevitably, some of the quite
compact structure of the previous subsection is lost, due to the sheer
number of possible pairs. Nevertheless, note that indeed members of
the same triplet $\{\F_1,\F_2,\F_3\}$, $\{\H_1,\H_2,\F_{3'}\}$ and
$\{\P_0,\P_3,\P_{3'}\}$ commute with each other. We defer
multiplication and anti-commutator tables to the appendix. Clearly the
nine generators are not unique. In the following section we will
relate a previously obtained third-rank tensor to a linear combination
of these generators. Before doing so we give explicit expressions for
the finite metamorphic transformations.

\begin{table}[h]
\begin{equation*}
\begin{array}{|c||ccc|ccc|ccc|}\hline
[\X,\Y]/2& \F_1& \F_2& \F_3& \H_1& \H_2& \F_{3'}& \P_0& \P_3& \P_{3'}\\
\hline\hline
\F_1& 0& 0& 0& q^2\B_{0'}& 0& -q^2\B_2& -\B_1& -q^2\D_2& 0\\
\F_2& 0& 0& 0& 0& q^4\B_0& -q^4\B_1& -\B_2& 0& -q^4\D_1\\
\F_3& 0& 0& 0& -q^2\D_2& -q^4\D_1& 0& 0& -q^6\B_{0'}& -q^6\B_0\\
\hline
\H_1& -q^2\B_{0'}& 0& q^2\D_2& 0& 0& 0& -\D_1& 0& q^2\B_2\\
\H_2& 0& -q^4\B_0& q^4\D_1& 0& 0& 0& -\D_2& q^4\B_1& 0\\
\F_{3'}& q^2\B_2& q^4\B_1& 0& 0& 0& 0& 0& -q^6\B_0& -q^6B_{0'}\\
\hline
\P_0& \B_1& \B_2& 0& \D_1& \D_2& 0& 0& 0& 0\\
\P_3& q^2\D_2& 0& q^6\B_{0'}& 0& -q^4\B_1& q^6\B_0& 0& 0& 0\\
\P_{3'}& 0& q^4\D_1& q^6\B_0& -q^2\B_2& 0& q^6\B_{0'}& 0& 0& 0\\
\hline
\hline
\B_0& 0& \H_2& \P_3'& 0& \F_2& \P_3& 0& \F_3'& \F_3\\
\B_2& -\F_3'& -\P_0 q^4& 0& -\P_3'& 0& -\F_1 q^4& -\F_2& 0& -\H_1 q^4\\
\D_2& -\P_3& 0& \H_1 q^4& -\F_3& \P_0 q^4& 0& -\H_2& \F_1 q^4& 0\\\hline
\B_0'& \H_1& 0& \P_3& \F_1& 0& \P_3'& 0& \F_3& \F_3'\\
\B_1& -\P_0 q^2& -\F_3'& 0& 0& -\P_3& -\F_2 q^2& -\F_1& -\H_2 q^2& 0\\
\D_1& 0& -\P_3'& \H_2 q^2& \P_0 q^2& -\F_3& 0& -\H_1& 0& \F_2 q^2\\\hline
\end{array}
\end{equation*}
\caption{Commutator relations $[\X,\Y]/2$ for the generators of
  metamorphic and isometric operations. $\X$ denotes a matrix of the
  leftmost column, $\Y$ one of the top row. The upper block of nine
  rows give the commutator between pairs of metamorphic
  generators. The lower block with 6 rows give the commutators
  betweenone isometric and one metamorphic generator.}
\label{TABcommutatorsOfMetamorphisms}
\end{table}

For the first triplet of generators, these are
\begin{align}
\exp(\tau_1\F_1)=&  \left(\begin{matrix}
    \cos(\tau_1q) & -q\sin(\tau_1q) & 0 & 0\\
    q^{-1}\sin(\tau_1q) & \cos(\tau_1q) & 0 & 0\\
    0 & 0 & \cos(\tau_1q) & -q\sin(\tau_1q)\\
    0 & 0 & q^{-1}\sin(\tau_1q) & \cos(\tau_1q)
  \end{matrix}
  \right),\\
\exp(\tau_2\F_2)=&  \left(\begin{matrix}
    \cos(\tau_2q^2) & 0 & -q^2\sin(\tau_2q^2) & 0\\
    0 & \cos(\tau_2q^2) & 0 & -q^2\sin(\tau_2q^2)\\
    q^{-2}\sin(\tau_2q^2) & 0 & \cos(\tau_2q^2) & 0\\
    0 & q^{-2}\sin(\tau_2q^2) & 0 & \cos(\tau_2q^2)
  \end{matrix}
  \right),\\
\exp(\tau_3\F_3)=&  \left(\begin{matrix}
    \cosh(\tau_3q^3) & 0 & 0 & -q^3\sinh(\tau_3q^3)\\
    0 & \cosh(\tau_3q^3) & q\sinh(\tau_3q^3) & 0\\
    0 & q^{-1}\sinh(\tau_3q^3) & \cosh(\tau_3q^3) & 0\\
    -q^{-3}\sinh(\tau_3q^3) & 0 & 0 & \cosh(\tau_3q^3)
  \end{matrix}
  \right).
\end{align}

The second group of finite metamorphic operations is
\begin{align}
\exp(\tau_1\H_1)=&  \left(\begin{matrix}
    \cosh(\tau_1q) & -q\sinh(\tau_1q) & 0 & 0\\
    -q^{-1}\sinh(\tau_1q) & \cosh(\tau_1q) & 0 & 0\\
    0 & 0 & \cosh(\tau_1q) & -q\sinh(\tau_1q)\\
    0 & 0 & -q^{-1}\sinh(\tau_1q) & \cosh(\tau_1q)
  \end{matrix}
  \right),\\
\exp(\tau_2\H_2)=&  \left(\begin{matrix}
    \cosh(\tau_2q^2) & 0 & -q^2\sinh(\tau_2q^2) & 0\\
    0 & \cosh(\tau_2q^2) & 0 & -q^2\sinh(\tau_2q^2)\\
    -q^{-2}\sinh(\tau_2q^2) & 0 & \cosh(\tau_2q^2) & 0\\
    0 & -q^{-2}\sinh(\tau_2q^2) & 0 & \cosh(\tau_2q^2)
  \end{matrix}
  \right),\\
\exp(\tau_3F_{3'})=&  \left(\begin{matrix}
    \cosh(\tau_3q^3) & 0 & 0 & -q^3\sinh(\tau_3q^3)\\
    0 & \cosh(\tau_3q^3) & -q\sinh(\tau_3q^3) & 0\\
    0 & -q^{-1}\sinh(\tau_3q^3) & \cosh(\tau_3q^3) & 0\\
    -q^{-3}\sinh(\tau_3q^3) & 0 & 0 & \cosh(\tau_3q^3)
  \end{matrix}
  \right).
\end{align}

And the third group of finite metamorphic operations is
\begin{align}
\exp(\tau_0\P_0)=&  \left(\begin{matrix}
    e^{\tau_0} & 0 & 0 & 0\\
    0 & e^{-\tau_0} & 0 & 0\\
    0 & 0 & e^{-\tau_0} & 0\\
    0 & 0 & 0 & e^{\tau_0}
  \end{matrix}
  \right),\\
\exp(\tau_3\P_3)=&  \left(\begin{matrix}
    \cos(\tau_3q^3) & 0 & 0 & -q^3\sin(\tau_3q^3)\\
    0 & \cos(\tau_3q^3) & -q\sin(\tau_3q^3) & 0\\
    0 & q^{-1}\sin(\tau_3q^3) & \cos(\tau_3q^3) & 0\\
    q^{-3}\sin(\tau_3q^3) & 0 & 0 & \cos(\tau_3q^3)
  \end{matrix}
  \right),\\
\exp(\tau_3\P_{3'})=&  \left(\begin{matrix}
    \cos(\tau_3q^3) & 0 & 0 & -q^3\sin(\tau_3q^3)\\
    0 & \cos(\tau_3q^3) & q\sin(\tau_3q^3) & 0\\
    0 & -q^{-1}\sin(\tau_3q^3) & \cos(\tau_3q^3) & 0\\
    q^{-3}\sin(\tau_3q^3) & 0 & 0 & \cos(\tau_3q^3)
  \end{matrix}
  \right).
\end{align}

\subsection{Morphological shifting and Jeffrey's third-rank tensor}
\label{sec:translations}

Based on the mathematical structure of Ref.\ \cite{schmidt04nahs}, in
Ref.\ \cite{schmidt07nage} a ``shifting operation'' was investigated
that changes the radius of a sphere by a given amount $R$. These
operations build a one-dimensional Abelian group. The generator of the
group, $\T_1$ (referred to as $\tilde {\bf G}$ in
\cite{schmidt07nage}), generates the kernel ${\sf K}_R$ of
Ref.~\cite{schmidt04nahs} upon exponentiation, ${\sf
  K}_R=\exp(RG)\equiv \exp(R\T_1)$. The significance of the matrix
${\sf K}_R$ lies i) in the algebraic structure: ${\sf K}_R\cdot {\sf
  K}_{R'}={\sf K}_{R'}\cdot{\sf K}_R={\sf K}_{R+R'}$, and ii) in the
fact that it contains the expressions for the four Kierlik-Rosinberg
weight functions explicitly.

Jeffrey's third-rank tensor as a central object of
Ref.~\cite{schmidt07nage} is given by the following set of four
matrices:
\begin{align}
 \T_0 &= \left(\begin{matrix}
    1 & 0 & 0 & 0\\
    0 & 1 & 0 & 0\\
    0 & 0 & 1 & 0\\
    0 & 0 & 0 & 1
  \end{matrix}\right),\quad
 \T_1 = \left(\begin{matrix}
    0 & 0 & 0 & -q^4/(8\pi)\\
    1 & 0 & -q^2/(4\pi) & 0\\
    0 & 8\pi & 0 & 0\\
    0 & 0 & 1 & 0
  \end{matrix}\right),\\
 \T_2 &= \left(\begin{matrix}
    0 & 0 & -q^4/(64\pi^2) & 0\\
    0 & -q^2/(4\pi) & 0 & -q^4/(64\pi^2)\\
    1 & 0 & -q^2/(4\pi) & 0\\
    0 & 1 & 0 & 0
  \end{matrix}\right),\\
 \T_3 &= \left(\begin{matrix}
    0 & -q^4/(8\pi) & 0 & -q^6/(32\pi^2)\\
    0 & 0 & -q^4/(64\pi^2) & 0\\
    0 & 0 & 0 & -q^4/(8\pi)\\
    1 & 0 & 0 & 0
  \end{matrix}\right),
\end{align}
where $\T_2=\T_1\cdot\T_1/(8\pi)$, $\T_3=-q^4(\T_1)^{-1}/(8\pi)$,
where $(\T_1)^{-1}$ is the inverse of $\T_1$. All $\T_\mu$ commute
with each other, $[\T_\mu,\T_\nu]=0$. Finite transformations are
obtained via exponentiation as $\exp(\chi_\nu \T_\nu)$, where
$\chi_\nu$ is the transformation parameter. The finite transforms
commute with each other and they obey $\exp(\sum_\nu \chi_\nu \T_\nu)
\exp(\sum_\nu \chi_\nu' \T_\nu)= \exp(\sum_\nu \chi_\nu' \T_\nu)
\exp(\sum_\nu \chi_\nu \T_\nu)= \exp(\sum_\nu(\chi_\nu+\chi_\nu')
\T_\nu)$.

\begin{table}[h]
\begin{equation*}
\begin{array}{|c||cccc|}\hline
[\X,\Y]/2& \T_0 & \T_1 & \T_2 & \T_3 \\
\hline\hline
  \T_0 & 0 & 0 & 0 & 0\\
  \T_1 & 0 & 0 & 0 & 0\\
  \T_2 & 0 & 0 & 0 & 0\\
  \T_3 & 0 & 0 & 0 & 0\\\hline
\end{array}
\quad
\begin{array}{|c||cccc|}\hline
\X\cdot\Y& \T_0 & \T_1 & \T_2 & \T_3 \\
\hline\hline
  \T_0 & \T_0 & \T_1 & \T_2 & \T_3\\
  \T_1 & \T_1 & 8\pi\T_2 & -\frac{q^2}{4\pi}\T_1+\T_3 & -\frac{q^4}{8\pi}\T_0\\
  \T_2 & \T_2 & -\frac{q^2}{4\pi}\T_1+\T_3 & -\frac{q^4}{64\pi^2}\T_0-\frac{q^2}{4\pi}\T_2 & -\frac{q^4}{64\pi^2}\T_1\\
  \T_3 & \T_3 & -\frac{q^4}{8\pi}\T_0 & -\frac{q^4}{64\pi^2}\T_1 & -\frac{q^6}{32\pi^2}\T_0-\frac{q^4}{8\pi}\T_2\\\hline
\end{array}
\end{equation*}
\caption{Tables for commutator relationships $[{\sf X},{\sf Y}]/2$ and
  for products $\X\cdot\Y$ for the generators of $\T_\mu$
  transformations.}
\label{tab:tAlgebra}
\end{table}

Explicitly, the matrices for finite transformations are given by
\begin{align}
  \exp(\chi_0\T_0)&=\left(\begin{matrix}
    {\rm e}^{\chi_0} & 0 & 0 & 0\\
    0 & {\rm e}^{\chi_0} & 0 & 0\\
    0 & 0 & {\rm e}^{\chi_0} & 0\\
    0 & 0 & 0 & {\rm e}^{\chi_0}
  \end{matrix}\right),\\
  \exp(\chi_1\T_1)&=\left(\begin{matrix}
      c+qs\chi_1/2 & (cq^2\chi_1-qs)/2 & -q^3 s \chi_1/(16\pi) & (cq^4\chi_1-3sq^3)/(16\pi)\\
      (s+cq\chi_1)/(2q) & c-(qs\chi_1)/2 & -(3sq+cq^2\chi_1)/(16\pi) & -q^3s\chi_1/(16\pi)\\
      4\pi s\chi_1/q & 4\pi(s+cq\chi_1)/q & c-(qs\chi_1)/2 & (cq^2\chi_1-sq)/2\\
      4\pi(s-cq\chi_1)/q^3 & 4\pi s\chi_1/q & (s+cq\chi_1)/(2q) & c+(qs\chi_1)/2
   \end{matrix}\right),\label{eqKmatrix}\\
  \exp(\chi_2\T_2)&=\left(\begin{matrix}
      g+(gq^2\chi_2)/(8\pi) & 0 & -gq^4\chi_2/(64\pi^2) & 0\\
      0 & g-gq^2\chi_2/(8\pi) & 0 & -gq^4\chi_2/(64\pi^2)\\
      g\chi_2 & 0 & g-(gq^2\chi_2)/(8\pi) & 0\\
      0 & g\chi_2 & 0 & g+(gq^2\chi_2)/(8\pi)
   \end{matrix}\right),\\
  \exp(\chi_3\T_3)&=
\left(\begin{matrix}
      C-(q^3S\chi_3)/(16\pi) & -(8\pi qS+Cq^4\chi_3)/(16\pi) &\\
      S/(2q)-(Cq^2\chi_3)/(16\pi) & C+(q^3S\chi_3)/(16\pi) &\\
      -qS\chi_3/2 & 4\pi S/q-(Cq^2\chi_3)/2 &\\
      4\pi S/q^3 +C\chi_3/2 & -qS\chi_3/2 &
   \end{matrix}\right.\nonumber\\
  &\hspace{30mm}\left.\begin{matrix}
  q^5S\chi_3/(128\pi^2) & -(24\pi q^3 S+Cq^6\chi_3)/(128\pi^2)\\
  (-24\pi qS+Cq^4\chi_3)/(128\pi^2) & q^5S\chi_3/(128\pi^2)\\
  C+(q^3S\chi_3)/(16\pi) & -(8\pi qS+Cq^4\chi_3)/(16\pi)\\
  S/(2q)-Cq^2\chi_3/(16\pi) & C-(q^3S\chi_3)/(16\pi)
\end{matrix}\right),
\end{align}
where we have used the short-hand notation $s=\sin(q\chi_1),
c=\cos(q\chi_1)$, $g=\exp(-q^2\chi_2/(8\pi))$,
$C=\cos(q^3\chi_3/(8\pi)), S=\sin(q^3\chi_3/(8\pi))$.  Eq.~(\ref{eqKmatrix})
describes ${\sf K}_R$ when setting $\chi_1=R$.

It is an interesting application of the theory outlined in the
previous section to try an express the $\T_\nu$ as linear combinations
of the metamorphic generators. This can indeed be done with a little
algebra, yielding the result:
\begin{align}
 \T_0&=\1,\\
 \T_1&= \frac{\F_1-\H_1}{2}
   +\frac{2\pi(\P_3-\P_{3'}+\F_3-\F_{3'})}{q^2}
   +\frac{3\P_3-\P_{3'}-\F_3+3\F_{3'}}{32\pi q^2},\\
 \T_2&=\frac{q^2(\P_0-\1)}{8\pi}
   +\frac{\F_2-\H_2}{2}+\frac{\F_2+\H_2}{128\pi^2},\\
 \T_3&=\frac{q^2(\F_1+\H_1)}{16\pi}
   +\frac{\P_3+\P_{3'}-\F_3-\F_{3'}}{4}
   +\frac{3\P_3+\P_{3'}+\F_3+3\F_{3'}}{256\pi^2}.
\end{align}

\section{Conclusions}
\label{sec:conclusions}

In conclusions we have presented a framework for manipulating
four-dimensional vector fields $\sf u$ that are defined on an
underlying three-dimensional Euclidian space. In real space, the
relevant operations are application of the Laplace operator and
building convolutions. These operations turn to multiplication by
$-q^2$ and the product operation in Fourier space. We have analysed
the symmetries that leave the metric (\ref{eq:metric}) for the
four-vectors invariant. This leads to operations that either leave the
metric invariant (isometric transforms) or that change the metric and
hence the morphology that the four-vectors describe (metamorphic
operations). We have kept the nature of the four-vectors general,
i.e.\ these can taken on aribtrary real values. This includes specific
geometries (such as spheres considered in Ref.\ \cite{schmidt07nage}),
but is more general. Whether the transformations presented here help
to contruct novel DFT approximations is an interesting question for
future work. It would also be interesting to explore possible
connections to the integral geometric framework by Hansen-Goos and
Mecke \cite{hansengoos}.

\subsection*{Acknowledgments} 

{\em The published version has a dedication to a very eminent
  theoretical physicist in it (guess who!).}
M R Dennis, G Leithall, G Rein, and F Catanese are acknowledged for
discussions, and M Burgis for a very careful reading of the
manuscript. This work was supported by the EPSRC under grant
EP/E065619/1 and by the SFB840/A3 of the DFG.

\clearpage
\appendix 
\section{Mixed commutator relations of isometric and metamorphic generators}

Here we give further details about the algebra of commutator
relations. Tab.\ \ref{TABmetamorphicProducts} gives a multiplication
table between metamorphic generators as well as anti-commutator
relationships. Tab.\ \ref{TABmetamorphicAndIsometricMixed} gives
products and anti-commutators between mixed pairs of an isometric and
a morphometric generator.

\begin{table}[h]
\begin{equation*}
\begin{array}{|c||ccc|ccc|ccc|}\hline
\X\cdot\Y& \F_1& \F_2& \F_3& \H_1& \H_2& \F_{3'}& \P_0& \P_3& \P_{3'}\\
\hline\hline
\F_1& -q^2\1& -F_3& q^2\F_2& q^2\B_{0'}& -\P_{3'}& -q^2\B_2& -\B_1& -q^2\D_2& q^2\H_2\\
\F_2& -\F_3& -q^4\1& q^4\F_1& -\P_3& q^4\B_0& -q^4\B_1& -\B_2& q^4\H_1& -q^4\D_1\\
\F_3& q^2\F_2& q^4\F_1& q^6\1& -q^2\D_2& -q^4\D_1& q^6\P_0& \F_{3'}& -q^6\B_{0'}& -q^6\B_0\\
\hline
\H_1& -q^2\B_{0'}& -\P_3& q^2\D_2& q^2\1& -\F_{3'}& -q^2\H_2& -\D_1& -q^2\F_2& q^2\B_2\\
\H_2& -\P_{3'}& -q^4\B_0& q^4\D_1& -\F_{3'}& q^4\1& -q^4\H_1& -\D_2& q^4\B_1& -q^4\F_1\\
\F_{3'}& q^2\B_2& q^4\B_1& q^6\P_0& -q^2\H_2& -q^4\H_1& q^6\1&\F_3& -q^6\B_0& -q^6B_{0'}\\
\hline
\P_0& \B_1& \B_2& \F_{3'}& \D_1& \D_2& \F_3& \1& \P_{3'}& \P_3\\
\P_3& q^2\D_2& q^4\H_1& q^6\B_{0'}& -q^2\F_2& -q^4\B_1& q^6\B_0& \P_{3'}& -q^6\1& -q^6\P_0\\
\P_{3'}& q^2\H_2& q^4\D_1& q^6\B_0& -q^2\B_2& -q^4\F_1& q^6\B_{0'}& \P_3& -q^6\P_0& -q^6\1\\
\hline
\end{array}
\end{equation*}
\begin{equation*}
\begin{array}{|c||ccc|ccc|ccc|}\hline
\{\X,\Y\}/2& \F_1& \F_2& \F_3& \H_1& \H_2& \F_{3'}& \P_0& \P_3& \P_{3'}\\
\hline\hline
\F_1& -q^2\1& -F_3& q^2\F_2& 0 & -\P_{3'}& 0& 0& 0& q^2\H_2\\
\F_2& -\F_3& -q^4\1& q^4\F_1& -\P_3& 0& 0& 0& q^4\H_1& 0\\
\F_3& q^2\F_2& q^4\F_1& q^6\1& 0& 0& q^6\P_0& \F_{3'}& 0& 0\\
\hline
\H_1& 0& -\P_3& 0& q^2\1& -\F_{3'}& -q^2\H_2& 0& -q^2\F_2& 0\\
\H_2& -\P_{3'}& 0& 0& -\F_{3'}& q^4\1& -q^4\H_1& 0& 0& -q^4\F_1\\
\F_{3'}& 0& 0& q^6\P_0& -q^2\H_2& -q^4\H_1& q^6\1&\F_3& 0& 0\\
\hline
\P_0& 0& 0& \F_{3'}& 0& 0& \F_3& \1& \P_{3'}& \P_3\\
\P_3& 0& q^4\H_1& 0& -q^2\F_2& 0& 0& \P_{3'}& -q^6\1& -q^6\P_0\\
\P_{3'}& q^2\H_2& 0& 0& 0& -q^4\F_1& 0& \P_3& -q^6\P_0& -q^6\1\\
\hline
\end{array}
\end{equation*}
\caption{Top: Multiplication table $\X\cdot\Y$ for products of the
  generators of metamorphic operations. Bottom: Anti-commutator
  relations $\{\X,\Y\}/2=(\X\cdot\Y+\Y\cdot\X)/2$ for generators of
  metamorphic transformations. $\X$ denotes a matrix of the leftmost
  column, $\Y$ one of the top row.}
\label{TABmetamorphicProducts}
\end{table}

\begin{table}
\begin{equation*}
\begin{array}{|c||ccc|ccc|ccc|}\hline
\X\cdot\Y& \F_1& \F_2& \F_3& \H_1& \H_2& \F_3'& \P_0& \P_3& \P_3'\\\hline\hline
\B_0& \D_1& \H_2& \P_3'& \B_1& \F_2& \P_3& \B_0'& \F_3'& \F_3\\
\B_2& -\F_3'& -\P_0 q^4& \B_1 q^4& -\P_3'& \B_0' q^4& -\F_1 q^4& -\F_2& \D_1 q^4& -\H_1 q^4\\
\D_2& -\P_3& -\B_0' q^4& \H_1 q^4& -\F_3& \P_0 q^4& -\D_1 q^4& -\H_2& \F_1 q^4& -\B_1 q^4\\\hline
\B_0'& \H_1& \D_2& \P_3& \F_1& \B_2& \P_3'& \B_0& \F_3& \F_3'\\
\B_1& -\P_0 q^2& -\F_3'& \B_2 q^2& \B_0 q^2& -\P_3& -\F_2 q^2& -\F_1& -\H_2 q^2& \D_2 q^2\\
\D_1& -\B_0 q^2& -\P_3'& \H_2 q^2& \P_0 q^2& -\F_3& -\D_2 q^2& -\H_1& -\B_2 q^2& \F_2 q^2\\\hline
\end{array}
\end{equation*}
\begin{equation*}
\begin{array}{|c||ccc|ccc|ccc|}\hline
\Y\cdot\X& \F_1& \F_2& \F_3& \H_1& \H_2& \F_3'& \P_0& \P_3& \P_3'\\\hline\hline
\B_0&\D_1& -\H_2& -\P_3'& \B_1& -\F_2& -\P_3& \B_0'& -\F_3'& -\F_3\\
\B_2&\F_3'& \P_0 q^4& \B_1 q^4& \P_3'& \B_0' q^4& \F_1 q^4& \F_2& \D_1 q^4& \H_1 q^4\\
\D_2&\P_3& -\B_0' q^4& -\H_1 q^4& \F_3& -\P_0 q^4& -\D_1 q^4& \H_2& -\F_1 q^4& -\B_1 q^4\\\hline
\B_0'&-\H_1& \D_2& -\P_3& -\F_1& \B_2& -\P_3'& \B_0& -\F_3& -\F_3'\\
\B_1& \P_0 q^2& \F_3'& \B_2 q^2& \B_0 q^2& \P_3& \F_2 q^2& \F_1& \H_2 q^2& \D_2 q^2\\
\D_1& -\B_0 q^2& \P_3'& -\H_2 q^2& -\P_0 q^2& \F_3& -\D_2 q^2& \H_1& -\B_2 q^2& -\F_2 q^2\\\hline
\end{array}
\end{equation*}
\begin{equation*}
\begin{array}{|c||ccc|ccc|ccc|}\hline
\{\X,\Y\}/2& \F_1& \F_2& \F_3& \H_1& \H_2& \F_3'& \P_0& \P_3& \P_3'\\\hline\hline
\B_0& \D_1& 0& 0& \B_1& 0& 0& \B_0'& 0& 0\\
\B_2& 0& 0& \B_1 q^4& 0& \B_0' q^4& 0& 0& \D_1 q^4& 0\\
\D_2& 0& -\B_0' q^4& 0& 0& 0& -\D_1 q^4& 0& 0& -\B_1 q^4\\\hline
\B_0'&0& \D_2& 0& 0& \B_2& 0& \B_0& 0& 0\\
\B_1& 0& 0& \B_2 q^2& \B_0 q^2& 0& 0& 0& 0& \D_2 q^2\\
\D_1& -\B_0 q^2& 0& 0& 0& 0& -\D_2 q^2& 0& -\B_2 q^2& 0\\\hline
\end{array}
\end{equation*}
\caption{Top: Multiplication table between generators of isomorphisms,
  $\X$, and generators of metamorphisms, $\Y$. Middle: Reverse product
  order, $\Y\cdot\X$. Bottom: Anti-commutator relationships for the
  same pairs.}
\label{TABmetamorphicAndIsometricMixed}
\end{table}

\clearpage

\end{document}